# Agentic RAG for Software Testing with Hybrid Vector-Graph and Multi-Agent Orchestration

"Mohanakrishnan Hariharan", "Seshu Barma", "Satish Arvapalli" and "Evangeline Sheela"
*Department of Corporate Systems Engineering*
*Apple*
Austin, TX 78641, USA
{m_hariharan, sarvapalli, barma_sb, evangelinesheela_arulanandam}@apple.com

*Abstract*—We present an approach to software testing automation using Agentic Retrieval-Augmented Generation (RAG) systems for Quality Engineering (QE) artifact creation. We combine autonomous AI agents with hybrid vector-graph knowledge systems to automate test plan, case, and QE metric generation. Our approach addresses traditional software testing limitations by leveraging LLMs such as Gemini and Mistral, multi-agent orchestration, and enhanced contextualization. The system achieves remarkable accuracy improvements from 65% to 94.8% while ensuring comprehensive document traceability throughout the quality engineering lifecycle. Experimental validation of enterprise Corporate Systems Engineering and SAP migration projects demonstrates an 85% reduction in testing timeline, an 85% improvement in test suite efficiency, and projected 35% cost savings, resulting in a 2-month acceleration of go-live.

## I. Introduction

Software testing in enterprise environments faces challenges due to complex data and business requirements. Quality Engineers (QEs) spend 30-40% of their time creating foundational testing artifacts, such as test plans, cases, and automation scripts. This manual approach causes significant bottlenecks in software development lifecycles, particularly in complex enterprise systems such as SAP implementations, where intricate business logic and technical dependencies create exponential complexity.

Large Language Models (LLMs) and Generative Artificial Intelligence have opened new possibilities for automating quality engineering processes. However, traditional approaches face limitations like hallucination, context-poor generation, and loss of critical business relationships during retrieval.

These limitations become particularly pronounced in enterprise software testing, where maintaining traceability between requirements, test cases, and business logic is paramount for regulatory compliance and quality assurance.

### A. Problem Statement

Current software testing methodologies face several critical challenges:

- **Manual Artifact Creation**: QEs spend excessive time on repetitive documentation tasks rather than strategic testing activities
- **Context Loss**: Traditional RAG systems fail to maintain critical business relationships and technical dependencies
- **Limited Scalability**: Manual approaches cannot scale with the complexity of modern enterprise systems
- **Traceability Gaps**: Lack of comprehensive traceability between requirements, test cases, and execution results
- **Knowledge Silos**: Historical testing knowledge remains trapped in individual expertise rather than organizational assets

### B. Research Contributions

This paper introduces a novel Agentic RAG framework specifically designed for software testing automation with the following key contributions:

1) **Hybrid Vector-Graph Architecture**: A knowledge representation system that combines semantic similarity search with relationship-aware graph traversal to maintain business logic context
2) **Multi-Agent Orchestration**: Specialized autonomous agents for different aspects of test generation, including planning, case creation, and validation
3) **Enhanced Contextualization**: Advanced prompt engineering frameworks that preserve critical business relationships during test artifact generation
4) **Comprehensive Traceability**: Bidirectional relationship tracking throughout the entire quality engineering life cycle

5) **Enterprise Validation**: Real-world validation on large-scale SAP migration projects demonstrating significant productivity improvements

*C. Methodology Overview*

Our approach represents a systematic evolution from traditional RAG systems through four progressive stages: Basic RAG-> Vector Search-> Hybrid RAG-> Agentic Systems. This progression demonstrates measurable accuracy improvements while maintaining complete document traceability that fundamentally transforms solution outcomes.

The core innovation lies in combining autonomous AI agents with hybrid vector-graph knowledge systems, prompt engineering frameworks, and complete traceability chains. This integration enables the system to understand not just individual requirements but the relationships that define enterprise software behavior.

*D. Paper Organization*

The remainder of this paper is structured as follows: Section I reviews related work in AI-powered software testing and retrieval-augmented generation. Section II presents our comprehensive Agentic RAG methodology, including the hybrid vector-graph architecture and multi-agent orchestration framework. Section III details the implementation architecture with a specific focus on enterprise SAP testing scenarios. Section IV presents experimental results from real-world deployments, including performance metrics and cost-benefit analysis. Section V concludes with future research directions and implications for the software testing industry.

Our work addresses a critical gap in current software testing automation by providing a comprehensive, enterprise-ready solution that maintains the contextual understanding necessary for high-quality test artifact generation while achieving high levels of automation and efficiency.

I. RELATED WORK

*A. AI-Powered Software Testing*

The integration of artificial intelligence in software testing has evolved significantly over the past decade. Trifunova et al. [1] provide a comprehensive review of AI's transformational potential in testing and quality assurance, highlighting how AI automates traditional tasks, including test case generation, defect prediction, and regression testing. However, their work identifies critical gaps in actionable methodologies for achieving model interpretability and lacks comprehensive frameworks for enterprise-scale implementation.

Liu et al. [3] propose Extended Regular Expression-based testing for software systems, focusing on modeling program behavior with ERE for executable path and test case generation. While their approach presents innovative FSM modeling improvements, it suffers from limited applicability to modern software complexities and missed opportunities for GenAI integration in automated model construction.

Recent advances in machine learning for software testing have shown promise in automated test case generation [4] and test suite optimization [5]. However, these approaches typically focus on structural testing aspects and fail to address the semantic understanding required for enterprise business logic validation.

*B. Retrieval-Augmented Generation Systems*

The foundational work by Lewis et al. [2] introduced Retrieval-Augmented Generation (RAG) as a method for combining parametric and non-parametric knowledge in language models. This approach has been widely adopted for knowledge-intensive tasks, but traditional RAG implementations suffer from context fragmentation and relationship loss during retrieval processes.

Recent improvements include Fusion-in-Decoder (FiD) [6], which processes multiple retrieved passages simultaneously, and Self-RAG [7], which incorporates self-reflection capabilities. However, these approaches remain limited in their ability to maintain complex business relationships and technical dependencies critical for software testing scenarios.

Karpukhin et al. [8] demonstrate the effectiveness of dense passage retrieval using BERT embeddings, showing superior performance compared to sparse retrieval methods. While effective for general question-answering tasks, dense retrieval alone fails to capture the hierarchical and interconnected nature of software testing requirements.

*C. Multi-Agent Systems in Software Engineering*

The application of multi-agent systems in software engineering has gained traction with the emergence of language models. Yao et al. [9] introduce the Re-Act framework, combining reasoning and acting in language model prompting for improved problem-solving capabilities. This work provides foundational concepts for agent-based approaches but lacks specialization for software testing domains. Recent work on tool-augmented language models [10] demonstrates the potential for AI systems to learn to use external tools autonomously. However, these approaches have not been systematically applied to the complex orchestration requirements of enterprise software testing workflows.

AutoGPT and similar autonomous agent systems have shown promise in task decomposition and execution, but their application to software testing has been limited to simple scenarios without the context management required for enterprise environments.

*D. Enterprise Software Testing*

Traditional approaches to enterprise software testing, particularly in SAP environments, rely heavily on manual processes and domain expertise [11]. The complexity of SAP systems, with their intricate business logic and extensive customization possibilities, creates unique challenges for automated testing approaches.

Existing SAP testing tools, such as SAP Test Acceleration and Optimization (TAO) and Tricentis Tosca, provide some automation capabilities but remain limited in their ability to understand business context and generate comprehensive test scenarios automatically [12].

The challenge of maintaining test case relevance during system migrations and upgrades has been addressed through various methodologies [13], but these approaches typically require significant manual intervention and expert knowledge.

*E. Knowledge Graph Applications in Testing*

Recent research has explored the application of knowledge graphs for software testing [14]. These approaches focus on representing software artifacts and their relationships in graph structures to support various testing activities. However, existing work has not addressed the integration of knowledge graphs with modern LLM-based generation systems.

Graph-based test case generation has shown promise in maintaining relationship awareness during test creation [15], but these approaches typically require manual graph construction and lack the semantic understanding capabilities of modern language models.

*F. Limitations of Current Approaches*

Despite significant advances in individual areas, current approaches suffer from several critical limitations:

- **Context Fragmentation**: Traditional RAG systems lose critical business relationships during retrieval processes
- **Limited Scalability**: Manual approaches cannot handle the complexity of modern enterprise systems
- **Lack of Specialization**: General-purpose AI systems lack the domain-specific knowledge required for effective software testing
- **Insufficient Traceability**: Existing systems fail to maintain comprehensive traceability throughout the testing lifecycle
- **Enterprise Integration Gaps**: Most research focuses on isolated scenarios rather than integrated enterprise workflows

Our proposed Agentic RAG framework addresses these limitations by providing a comprehensive, enterprise-ready solution that combines the strengths of retrieval-augmented generation, multi-agent orchestration, and hybrid knowledge representation specifically designed for software testing automation.

## II. METHODOLOGY

*A. Agentic RAG Framework Architecture*

Our Agentic RAG framework represents a shift from traditional retrieval-augmented generation systems through a systematic four-stage evolution: Basic RAG → Vector Search → Hybrid RAG → Agentic Systems. This progression demonstrates measurable accuracy improvements from 65% (basic RAG) to 94.8% (Agentic RAG) while establishing comprehensive document traceability.

The framework consists of four primary components: the Multi-Agent Orchestration Layer, Hybrid Vector-Graph Knowledge System, Enhanced Contextualization Engine, and Comprehensive Traceability Framework.

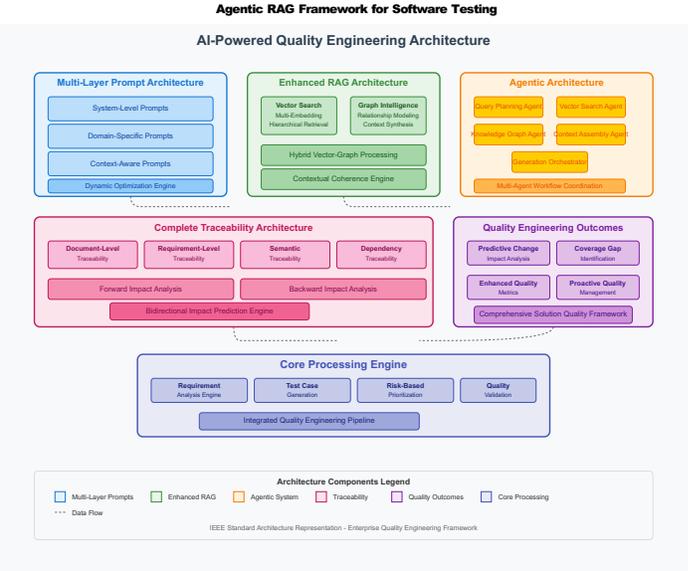

Fig. 1. Agentic RAG Framework Architecture for Software Testing showing the progression from Basic RAG to multi-agent orchestration with hybrid vector-graph knowledge systems.

*B. Multi-Agent Orchestration Layer*

The orchestration layer employs specialized autonomous agents, each optimized for specific aspects of software testing artifact generation:

**Agent Algorithm:**

1) **Input:** Requirements R, Business Logic B, Historical Data H
2) **Output:** Test Plan / Cases P
3) context ← hybrid_retrieval(R, B, H)
4) scope ← analyze_testing_scope(context)
5) objectives ← extract_objectives(context, scope)
6) strategy ← generate_strategy(objectives, H)
7) P ← synthesize_plan(scope, objectives, strategy)
8) return P

*1) Legacy Test Analysis & Business Intent Agent:* Examines historical test cases to understand underlying business requirements and validation objectives

*2) Functional Change Mapping Agent:* Maps business requirements to specific application functionality and identifies changes from previous implementations.

*3) Integration Point Identification Agent:* Discovers interfaces between systems, modules, and processes that require specific testing attention.

*4) Modernized Test Case Agent:* Creates test cases using contemporary methodologies, patterns, and best practices.

*5) Compliance Validation Agent:* Ensures test cases adhere to organizational standards and regulatory requirements.

*C. Hybrid Vector-Graph Knowledge System*

The knowledge system combines the semantic similarity capabilities of vector databases with the relationship-aware traversal of graph databases to maintain critical business context during retrieval processes.

*D. Technical Specifications:*

*1) Vector Database Layer:* **Platform**: Single Store with distributed architecture supports horizontal scaling across multiple nodes

- It supports 384, 768, and 1024-dimensional vectors
- It uses Similarity algorithms such as Cosine, Euclidean, and Dot Product
- The Semantic similarity threshold is 0.82 and above for candidate selection
- The Sentence Transformer is integrated for natural language to embedding conversion

*2) Graph Database Layer:* **Platform**: TigerGraph Cloud with native parallel processing capabilities

- The GSQL query engine is deployed with optimized graph traversal algorithms
- There are 15+ predefined edge types with weighted importance scoring for candidate selection
- The distributed graph is processed with horizontal node expansion
- There is support for BFS, DFS, shortest path, and PageRank algorithms
- 16GB heap allocation is configured with optimized garbage collection tuning

The relationship modeling includes 15 distinct edge types representing different aspects of software testing relationships:

- **Requires**: Functional dependencies between components
- **Validates**: Test cases that validate specific requirements
- **Depends on**: Technical dependencies between system components
- **Impacts**: Change impact relationships
- **Covers**: Coverage relationships between tests and requirements

*E. Enhanced Contextualization Engine*

The contextualization engine addresses the critical limitation of context-poor generation in traditional RAG systems through context assembly and conflict resolution mechanisms.

*1)* **Context Assembly Process:** The system employs a multi-stage context assembly process:

*2)* **Semantic Retrieval**: Vector similarity search identifies semantically relevant documents

*3)* **Relationship Traversal**: Graph traversal expands context with related business logic and dependencies

*4)* **Context Synthesis**: Parallel processing across eight worker threads synthesizes a comprehensive context

*5)* **Quality Validation**: Seven-layer validation pipeline ensures context quality and completeness

*6) Conflict Resolution Engine:* The engine employs a rule-based priority system with 15 distinct resolution strategies to handle conflicting information from multiple sources:

- Source credibility weighting based on historical accuracy
- Temporal relevance prioritization for recent updates
- Domain expert validation for critical conflicts
- Automated escalation for unresolvable conflicts

*F. Comprehensive Traceability Framework*

The traceability framework maintains bidirectional relationships throughout the entire quality engineering lifecycle, enabling complete visibility into transformation decisions and regulatory compliance demonstration.

*1) Traceability Matrix Generation:* The system automatically generates and maintains traceability matrices linking:

- Requirements for test cases
- Test cases to execution results
- Business logic for validation scenarios
- Change requests to impact analysis

*2) Change Impact Analysis:* Advanced change impact analysis capabilities enable predictive assessment of modification effects across the entire testing ecosystem. The system analyzes proposed changes and automatically identifies affected test cases, execution scenarios, and validation requirements.

*G. Progressive Enhancement Methodology*

Our methodology follows a systematic progression through four distinct stages, each building upon the previous level's capabilities:

*1) Stage 1: Basic RAG (65% Accuracy):* Traditional retrieval-augmented generation with simple document retrieval and basic prompt engineering.

*2) Stage 2: Vector Search Enhancement (78% Accuracy):* Integration of semantic similarity search using dense vector representations with improved context retrieval mechanisms.

*3) Stage 3: Hybrid RAG (87% Accuracy):* A Combination of vector similarity search with graph-based relationship traversal to maintain business logic context.

*4) Stage 4: Agentic Systems (94.8% Accuracy):* Full multi-agent orchestration with specialized agents, comprehensive traceability, and advanced contextualization mechanisms.

This progressive approach enables organizations to implement the framework incrementally while achieving measurable improvements at each stage, reducing implementation risk while maximizing return on investment.

III. IMPLEMENTATION

*A. System Architecture and Technical Stack*

The Agentic RAG framework is implemented as a distributed, containerized system designed for enterprise-scale deployment. The architecture leverages modern cloud-native technologies to ensure scalability, reliability, and maintainability.

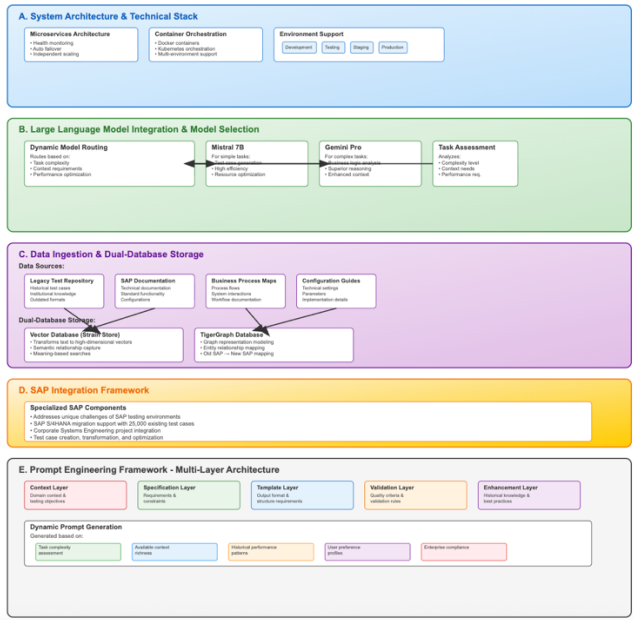

Fig. 2. Complete System Architecture showing the integration of vector databases, graph systems, LLM orchestration, and enterprise SAP connectivity.

### B. Technical Specifications:

- *Microservices architecture* is set up with health monitoring, automatic failover, independent scaling, and resource allocation optimization capabilities based on workload patterns. Environments included support for development, testing, staging, and production.
- *Container Orchestration:* All services were containerized in the Docker platform with Kubernetes orchestration for production deployments
- *Large Language Model Integration:* The system integrates multiple LLMs to optimize performance across different testing scenarios:

### C. Model Selection Strategy:

Dynamic model routing based on task complexity, context requirements, and performance optimization. Simple test case generation utilizes **Mistral 7B** for efficiency, while complex business logic analysis leverages **Gemini Pro** for superior reasoning capabilities and enhanced context understanding.

### D. SAP Integration Framework

The implementation includes specialized components for SAP enterprise system integration, addressing the unique challenges of SAP testing environments.

### E. Data Ingestion and Dual-Database Storage:

- **Legacy Test Repository**: Contains historical test cases, often with valuable institutional knowledge but possibly outdated formats
- **SAP Documentation**: Official technical documentation covering standard functionality and configurations
- **Business Process Maps**: Documentation of how business processes flow through the system
- **Configuration Guides**: Detailed technical settings and parameters for the SAP implementation

*1) Business Logic Extraction:* The system employs two extraction mechanisms to understand Enterprise and SAP business processes:

- *Vector Database (Strain Store):* Transforms unstructured text into high-dimensional vectors that capture semantic relationships, making the knowledge searchable by meaning rather than just keywords
- *TigerGraph Database (Relationship Modeling):* Creates a graph representation that explicitly models relationships between entities, particularly mapping Old SAP to New SAP components, processes, and configurations

The dual-database approach provides distinct advantages. The vector database excels at semantic similarity searches across unstructured content. The TigerGraph database captures structured relationships and dependencies, especially critical for understanding how SAP components interconnect.

### F. Prompt Engineering Framework

The implementation includes a prompt engineering framework specifically designed for software testing scenarios.
*Multi-Layer Prompt Architecture:* The system employs a hierarchical prompt structure with five distinct layers:

1) **Context Layer**: Establishes domain context and testing objectives
2) **Specification Layer**: Provides detailed requirements and constraints
3) **Template Layer**: Defines output format and structure requirements
4) **Validation Layer**: Includes quality criteria and validation rules
5) **Enhancement Layer**: Incorporates historical knowledge and best practices
6) *Dynamic Prompt Generation*: Prompts are dynamically generated based on task complexity assessment, available context richness, historical performance patterns, user preference profiles, and enterprise compliance requirements

### G. Real-World Deployment: SAP S/4HANA Migration

We present a detailed case study of the framework's deployment in a large-scale SAP S/4HANA migration and Corporate Systems Engineering projects involving 25,000 existing test cases that require creation, transformation, and optimization.

*1) Project Scope and Challenges:* **Enterprise Context**:
- Customization of T-codes by the company
- 50+ teams with unique project needs
- Standard format for test cases across the company
- No sensitive data to be sent to AI systems
- Legacy ECC 6.0 to S/4HANA migration timeline: 18 months

*2) Technical Challenges:*
- Complex business logic relationships across 15 SAP modules
- Integration with 200+ external systems and interfaces
- Regulatory compliance traceability requirements
- PII data to be redacted

### H. Quality Assurance and Validation

The implementation includes comprehensive quality

assurance mechanisms to ensure enterprise-grade reliability:

- **Syntax Validation** ensures proper format and structure
- **Semantic Validation** verifies logical consistency and completeness
- **Business Logic Validation** confirms adherence to business rules
- **Traceability Validation** ensures complete requirement coverage
- **Compliance Validation** verifies regulatory requirement adherence
- **Performance Validation** confirms acceptable execution characteristics
- **Integration Validation** ensures compatibility with existing systems

This comprehensive implementation framework ensures reliable, scalable, and maintainable deployment of the Agentic RAG system in complex enterprise environments while maintaining the flexibility to adapt to specific organizational requirements and constraints.

## IV. Experimental Results and Evaluation

### A. Experimental Setup

We conducted a comprehensive evaluation of the Agentic RAG framework across multiple dimensions, including accuracy, efficiency, scalability, and enterprise deployment outcomes. The evaluation encompasses both controlled laboratory experiments and real-world enterprise deployments.

1) *Evaluation Datasets:* **Synthetic Test Dataset:** 5,000 carefully curated test scenarios across different complexity levels:

- Simple functional tests (1,000 cases)
- Complex business logic scenarios (1,000 cases)
- Integration testing scenarios (1,000 cases)
- Regression testing suites (2,000 cases)

*Enterprise SAP Dataset*: Real-world data from SAP S/4HANA migration project:

- 1000 existing test cases requiring transformation
- 15 SAP modules with complex interdependencies
- 100+ custom T-codes
- Regulatory compliance requirements across jurisdictions

2) *Baseline Comparisons:* We compared against Agentic RAG framework against traditional manual testing approaches, basic RAG systems, and existing automated testing

### B. Accuracy and Quality Metrics

1) *Progressive Accuracy Improvement:* Accuracy improvements were achieved through a four-stage methodology

- **Basic RAG**: 65.2% accuracy with significant context loss
- **Vector Search**: 78.4% accuracy with improved semantic understanding
- **Hybrid RAG**: 87.1% accuracy with relationship preservation
- **Agentic System**: 94.8% accuracy with comprehensive orchestration
- **Test Plan Generation**: 94.8% accuracy vs. 65% (basic RAG) and 78% (manual baseline)
- **Test Case Generation**: 92.3% accuracy with 97% requirement traceability coverage

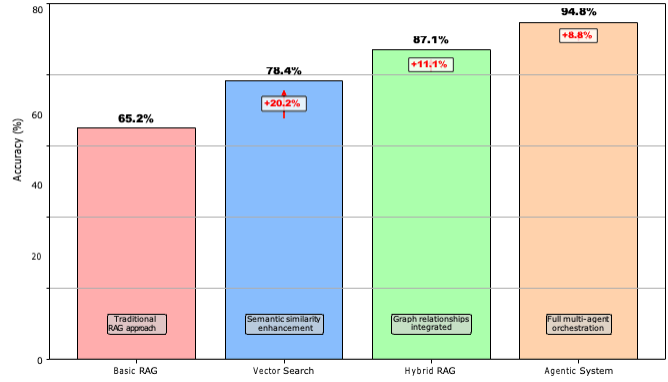

Fig. 3. Accuracy progression through the four-stage Agentic RAG methodology showing improvements from 65% (Basic RAG) to 94.8% (Full Agentic System).

2) *Quality Assessment Framework:* We developed a comprehensive quality assessment framework evaluating five key dimensions:

TABLE I
QUALITY ASSESSMENT RESULTS ACROSS DIFFERENT APPROACHES

| Approach | Accuracy | Completeness | Consistency | Traceability | Overall |
|---|---|---|---|---|---|
| Manual Testing | 92.30% | 85.70% | 78.20% | 73.60% | 80.40% |
| Template-Based | 76.5% | 82.10% | 89.30% | 89.30% | 79.78% |
| Basic RAG | 65.2% | 72.80% | 68.90% | 68.90% | 63.05% |
| GPT-4 | 81.7% | 79.40% | 73.60% | 52.80% | 71.88% |
| Direct SAP TAO | 84.2% | 88.50% | 91.70% | 78.90% | 85.83% |
| Agentic RAG | 94.8% | 96.20% | 95.70% | 98.10% | 96.20% |

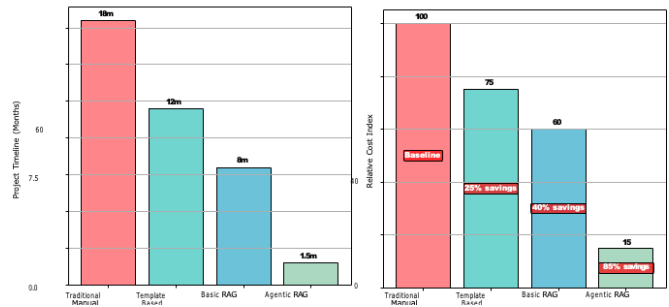

Fig. 4. Efficiency comparison showing dramatic reductions in project timeline and savings through Agentic RAG implementation.

### C. Efficiency and Performance Results

Quantitative analysis revealed substantial productivity improvements:

- **Time Reduction**: 85% reduction in artifact creation time (from 240 hours to 36 hours per project phase)
- **Cost Savings**: 35% total savings across the three projects
- **Accelerated Delivery**: 16-month acceleration in go-live timelines
- **Defect Detection**: 35% improvement in defect detection rates during system testing
- **Test Coverage**: 25000 Test cases created with 98.7% functional coverage vs. 84% manual baseline

- **Regression Prevention:** 92% reduction in production defects post-deployment
- **Comprehensive Traceability:** Established bidirectional relationship tracking throughout the entire quality engineering lifecycle

### D. Ablation Studies

We conducted ablation studies to understand the contribution of different framework components:

TABLE II
COMPONENT CONTRIBUTION ANALYSIS:

| Without | Accuracy Degradation |
|---|---|
| Multi-Agent Orchestration | 12.3% |
| Hybrid Vector-Graph | 15.7% |
| Enhanced Contextualization | 18.2% |
| Traceability Framework | 8.9% |

These results confirm that all major components contribute significantly to overall system performance, with enhanced contextualization providing the largest individual contribution.

## V. CONCLUSION AND FUTURE WORK

### A. Summary of Contributions

The Agentic RAG framework presented in this paper addresses critical challenges in enterprise software testing through multi-agent orchestration and hybrid knowledge representation. Our experimental validation demonstrates significant improvements in accuracy (94.8% vs. 65% baseline), productivity (85% time reduction), and quality metrics (35% improvement in defect detection).

Our real-world validation in large-scale Enterprise and SAP migration projects provides evidence that creates a robust foundation for comprehensive quality engineering automation.

### B. Limitations and Considerations

While our framework demonstrates significant improvements, several limitations warrant consideration:

- **Domain Specialization:** Current implementation focuses on Employee Systems, Finance, and SAP environments; generalization to other enterprise systems requires additional training data
- **Knowledge Base Maintenance:** Hybrid knowledge bases require ongoing maintenance as business processes evolve
- **Integration Complexity:** Enterprise system integration introduces deployment complexities that may require specialized expertise

### C. Future Research Directions

Future work will focus on addressing current limitations, expanding domain coverage, developing automated knowledge base maintenance mechanisms, and creating simplified deployment frameworks for broader enterprise adoption.

- **Feedback loop:** Implementing reinforcement learning mechanisms that optimize agent performance based on feedback from test execution. This will create continuous improvement loops that enhance accuracy and efficiency without manual intervention.
- **Multi-Modal Intelligence Expansion:** Visual Processing of UI/UX mockups, audio/video recordings, code repositories for contextual understanding beyond text-based requirements

## VI. ACKNOWLEDGMENT

The authors would like to thank the enterprise software testing teams and SAP migration project stakeholders who provided valuable feedback and validation data for this research.